\documentstyle[preprint,aps]{revtex}   
\begin{document}
\draft
\title{Magnetism of CaV$_{\rm 2}$O$_{\rm 5}$, CaV$_{\rm 3}$O$_{\rm 7}$,
CaV$_{\rm 4}$O$_{\rm 9}$: quantum effects or orbital ordering?}

\author{S. Marini and D. I. Khomskii} 

\address{Laboratory of Applied and Solid State Physics, 
Materials~Science~Centre, \\  
University~of~Groningen, 
Nijenborgh~4, 
9747~AG~Groningen,
The~Netherlands} 
\maketitle
\begin{abstract}

The quasi 2-d spin $\frac{1}{2}$ materials 
CaV$_{\rm n}$O$_{\rm 2n+1}$ (n=2,3,4) are often treated as systems in 
which quantum effects play a
dominant role: CaV$_{\rm 2}$O$_{\rm 5}$ has a spin gap and is thought 
to be a spin ladder; CaV$_{\rm 3}$O$_{\rm 7}$ has unusual long-range 
magnetic order which is explained by quantum fluctuations; 
the spin gap in CaV$_{\rm 4}$O$_{\rm 9}$ is usually attributed to
the formation of plaquette RVB. We show that in all these compounds there
should exist an orbital ordering which strongly modifies the exchange 
interaction and provides an alternative explanation of
 their magnetic properties, without invoking special quantum effects. 
 The type of magnetic ordering obtained for CaV$_{\rm 3}$O$_{\rm 7}$
quite naturally explains its magnetic structure which turns out to be 
that of a quasi 1-d antiferromagnet. The structure of  
CaV$_{\rm 4}$O$_{\rm 9}$ corresponds to singlet dimers rather than to a 
plaquette RVB. Singlet dimers should exist also in CaV$_{\rm 2}$O$_{\rm 5}$. 
The chirality of the crystal structure of CaV$_{\rm 4}$O$_{\rm 9}$ is also mentioned 
and some possible consequences are briefly discussed.
\\ 
\end{abstract}

\narrowtext

Low dimensional magnetic materials attract now considerable attention. Particularly
interesting are the systems with a marked quantum behaviour, having singlet
ground states with spin gaps (materials with spin-Peierls transitions, spin-ladders, 
layered cuprates etc.). Most of such systems studied up to now contain 
Cu$^{\rm 2+}$ ions, 
which  have one hole in the d-shell and consequently spin $\frac{1}{2}$.
Recently a new class of V compounds (CaV$_{\rm n}$O$_{\rm 2n+1}$,
n=2,3,4) was found with rather
unusual magnetic properties, mostly ascribed to quantum effects. They
contain V$^{\rm 4+}$ ions with one d-electron, so that they are 
often treated as electron counterparts of the cuprates. 
 The first member of this family, CaV$_{\rm 2}$O$_{\rm 5}$, was
found to have a spin gap and is considered as a spin ladder~\cite{Iwase96}. 
The second one, 
CaV$_{\rm 3}$O$_{\rm 7}$, has long range antiferromagnetic ordering at 
T$_{\rm N} \simeq $23 K with an unusual magnetic structure~\cite{Harashina96}.
Quasiclassical treatment of the Heisenberg model was not able 
to reproduce this structure, which stimulated the authors of Ref.~\cite{Kontani396} 
to invoke quantum fluctuations in order 
to explain it. And, last but not least, CaV$_{\rm 4}$O$_{\rm 9}$ was the 
first of these systems to attract considerable attention. Immediately after 
the first
experimental paper~\cite{Taniguchi95}, there was a burst of theoretical
activity around this compound~\cite{Ueda96}, which was treated as the first 
example of a new situation, the so 
called plaquette RVB --- a singlet state on 4 sites of a plaquette. 
 This is of course a very exciting development. We want to point 
out, however, that the actual situation in the system CaV$_{\rm n}$O$_{\rm 2n+1}$
may be quite different from what was discussed in 
Refs.~\cite{Iwase96}-~\cite{Ueda96}:
 the orbital degrees of freedom, completely ignored up to now,
play here a crucial role and provide a different explanation of the observed
phenomena. In the previous treatments~\cite{Iwase96}-~\cite{Ueda96} one proceeded 
from the homogeneus
Heisenberg interaction. In this model, the orbital character 
of the electronic wave function is ignored, and the exchange interaction is taken as
isotropic. However, if ions with degenerate orbital configurations are present, 
according to 
Jahn-Teller theorem the degeneracy must be lifted in the ground state:  
a particular orbital is stabilized at each site, 
making the exchange interaction inhomogeneus. This orbital ordering 
(cf.~\cite{Kugel73})
gives a very simple explanation of the magnetic properties of 
CaV$_{\rm n}$O$_{\rm 2n+1}$,
without invoking special quantum effects.\\
\hspace*{3mm} The 
fundamental structural unit of the CaV$_{\rm n}$O$_{\rm 2n+1}$ compounds 
is a VO$_{\rm 5}$ pyramid with a square base, 
containing the V ion slightly shifted from the basal plane. 
Alternated up and down pyramids line up, with their square bases 
sharing common edges and forming 2-d V-O sheets; Ca ions are located
between these sheets. In
different members of this family, certain V ions are missing in a regular 
fashion, so that we can describe the CaV$_{\rm n}$O$_{\rm 2n+1}$ series 
by a 
2-d square lattice with every third, fourth or fifth ion
missing. In a square 
pyramid coordination, d-states are split into a ground state doublet
(d$_{\rm xz}$, d$_{\rm yz}$), and a higher energy singlet 
d$_{\rm xy}$; this is confirmed by ESR (Taniguchi et al, 
cited in~\cite{Harashina96}).
The d-electron of V$^{\rm 4+}$ occupies at each site one of these
degenerate states or their linear combination: 
$|\alpha>$=cos$\alpha|$d$_{\rm xz}$$>$+sin$\alpha|$d$_{\rm yz}$$>$,
where the angle $\alpha$ gives the orientation of the plane in which the
orbital lies with respect to the x-axis. We will show that the crystal structure 
selects the value of $\alpha$,
leading in each compound to certain orbital 
ordering. The simple consideration that, due to this selective occupation of
orbitals, each V has different exchange coupling with different neighbours, 
explains the failure of an approach which uses an uniform Heisenberg interaction.
With the proper orbital ordering, the magnetic behaviour is straighforwardly 
determined by the 
Goodenough-Kanamori-Anderson (GKA) rules~\cite{Goodenough},
and the appeal to subtle quantum effects becomes unnecessary.\\
\hspace*{3mm} We start with CaV$_{\rm 3}$O$_{\rm 7}$, which turns out to be the 
simplest and clearest case. The basic elements of its crystal structure are shown 
in fig.1 together with
the orbital ordering obtained in~\cite{Harashina96}. The attempts to 
reproduce it on the basis of the homogeneus Heisenberg interaction in a mean 
field approximation failed
for any values of the nn exchange J$_{\rm e}$
(across the edges of plaquettes) and nnn exchange J$_{\rm c}$(across the corners)
~\cite{Kontani396}. 
As a possible way out, the inclusion of quantum fluctuations was
suggested in~\cite{Kontani396}, but agreement with the 
experimental magnetic
structure was found only in a narrow range of parameters. We show
now that this magnetic structure is a natural consequence of the 
orbital ordering which should exist in this compound. Two kinds of V sites 
are present in this
lattice: the one symmetrically surrounded by two nn V ions and
two nn empty plaquettes (site A), and the one with three nn V ions and 
one nn empty plaquette (sites B, B', C). The d-electron at a site A obviously 
prefers to occupy an orbital pointing towards the two positive charges in B 
and B', i.e. there should exist an extra crystal field splitting of the 
d$_{\rm xz}$ and 
d$_{\rm yz}$ orbitals stabilizing the d$_{\rm xz}$. For the same reasons, the 
d$_{\rm yz}$ orbital will be occupied at site B, B',C .
To check these arguments we carried out simple calculations treating the 
crystal field in a point charge model.
For the site A and in a nn approach, the electrostatic energy splitting we are 
interested in is given by   
\begin{eqnarray} \Delta_{\rm CF}= 2 \int \frac{- 4 \, e^{\rm 2} \, \left( \left| 
\psi_{\rm yz} (\bf{r'}) \right| ^{\rm 2} - \left| \psi_{\rm xz} 
(\bf{r'}) \right| ^{\rm 2} \right) \, 
d \bf{r'} }{ \left| \bf{r} - \bf{r'} \right| },
\end{eqnarray}
where $\bf{r}$ and $\bf {r'} $ are the positions of the V$^{\rm 4+}$
ion in B (or B') and of the d-electron respectively. A standard expansion 
of the 1/r point charge potential in spherical harmonics shows that only the
quadrupole terms contribute, giving
\begin{eqnarray} \Delta_{\rm CF}= (\bf{Q}_{\rm dxz}-\bf{Q}_{\rm dyz}) \bf{\cdot} 
\bf{\nabla E}_{\rm CF}
\end{eqnarray}
(contraction of the quadrupole tensor of the d orbitals $\bf{Q}$, and 
the electric field gradient (EFG) 
tensor $\bf{\nabla E}_{\rm CF}$ created
at the V site by the point-charge neighbours). 
The tensor $(\bf{Q}_{\rm dxz}-
\bf{Q}_{\rm dyz})$ is traceless and diagonal, its xx 
component being q=9.15$\cdot$10$^{\rm -28}$ in cgs units,
while the zz one is zero for symmetry reasons. We then proceed by calculating
and diagonalizing the EFG tensor, whose principal axes fix a new frame of reference
$x'y'z'$ in which the expression for the splitting becomes 
$\Delta_{\rm CF}$=q$\cdot (\bf{\nabla E}_{\rm CF}^{\rm x'x'}-
\bf{\nabla E}_{\rm CF}^{\rm y'y'})$ and the
occupied orbital is d$_{\rm x'z'}$ or d$_{\rm y'z'}$ 
according to the sign of $\Delta_{\rm CF}$. 
The resulting splitting at site A is
$\Delta_{CF} \simeq $0.36 eV, the
d$_{\rm xz}$ orbital being the lowest one. In a more detailed treatment 
with 20 neighbours $\Delta_{CF} \simeq $0.33 eV. Covalency contributions 
to the splitting play here a 
minor role because the t$_{\rm 2g}$ orbitals do not point towards the oxygen 
ligands, and they have been neglected. A more subtle point to consider is
the out of plane position of the V ions with respect to the oxygen lattice.
According to structural data~\cite{Bouloux73,Bouloux76},
the angle between the V-basal oxygen line and the basal plane is about 
25$^{\circ}$  in
all the CaV$_{\rm n}$O$_{\rm 2n+1}$ compounds. If this is taken into 
account in the EFG tensor, for the 20 neighbours cluster 
the splitting is reduced to about 0.31 eV, which is still large enough to
support our idea: up to reasonable temperatures, only the d$_{\rm xz}$ 
orbital is
occupied at site A. Similar treatment shows that the d$_{\rm yz}$ orbital 
is stabilized at site B, B' and C, with the stabilization energy 
$\Delta_{CF}$(B)=0.5 $\Delta_{CF}$(A), in agreement
with the symmetry considerations. The conclusion is that, simply due 
to the crystal structure, there exists in CaV$_{\rm 3}$O$_{\rm 7}$
the orbital ordering represented in fig.1, where the projection
of the occupied orbitals on the basal plane is shown.
This ordering should 
be accompanied by an average elongation of plaquettes along the y
direction, which actually has been detected~\cite{Bouloux73}.\\
\hspace*{3mm} From the GKA
rules~\cite{Goodenough} it is then evident that there will be strong antiferromagnetic
exchange along the y-direction,
whereas in the x-direction the interaction will be ferromagnetic and much
weaker. (Although nn V-O-V coupling along the antiferromagnetic chain is 
a 90$^{\circ}$ one, it is antiferromagnetic because we are 
dealing with the $\pi$-bonding of d$_{\rm yz}$ orbitals). This quite naturally 
explains the magnetic structure 
of fig.1. The nnn interaction across the corners of
the plaquettes, namely between sites A and C, will be 
antiferromagnetic too, once more
supporting the long range magnetic order experimentally observed. 
With this type of orbital ordering one can visualize the magnetic 
properties of CaV$_{\rm 3}$O$_{\rm 7}$ as those of weakly coupled 
1-d antiferromagnetic chains parallel to the y-axis (in this respect 
CaV$_{\rm 3}$O$_{\rm 7}$
is analogous to KCuF$_{\rm 3}$, which is a very good quasi 1-d 
antiferromagnet due to a particular type of orbital 
ordering~\cite{Kugel73}). This conclusion finds support in the behaviour of the 
magnetic susceptibility~\cite{Harashina96}, which, as typical for quasi 1-d 
antiferromagnets, 
has a pronounced maximum at T$_{\rm max} \simeq $95 K, much higher than 
T$_{\rm N} \simeq $23 K. \nopagebreak[3]We checked that the whole $\chi$(T) 
curve can be fitted rather well by the 
Bonner-Fisher expression~\cite{BonnerFisher} for temperatures above 
35 K. The standard treatment~\cite{BonnerFisher} gives that the exchange 
integral along the chain is J$_{\rm 1d} \simeq{T_{\rm max}}$$/$0.64 $\simeq$150 K. 
Weaker interactions between the chains give 3-d ordering at much 
lower temperature (T$_{\rm N} \simeq $23 K).\\
\hspace*{3mm} We turn now to the most interesting
case, CaV$_{\rm 4}$O$_{\rm 9}$. The spin gap 
observed in~\cite{Taniguchi95} was generally attributed to plaquette RVB~\cite{Ueda96}, 
but we show that the inclusion of orbital degrees of 
freedom again drastically modifies the situation. 
The schematic elements of the crystal structure of this compound~\cite{Bouloux73} 
are shown 
in fig.2. (This structure is rather interesting in itself: one can 
visualize it as the trace of the walk of the "left-moving knight" on a 
chess-board. Of course the right-hand variant exists as well. Correspondingly,
each plane has certain chirality, and the whole sample may be 
chiral too, displaying for example an optical activity and many other 
interesting properties). In CaV$_{\rm 4}$O$_{\rm 9}$ every V ion has the same
surrounding, i.e. just one kind of site exists in which the electric
field is rather asymmetric. As far as the 4 nn plaquettes of a V ion are
concerned, the situation is the same as at site B in
CaV$_{\rm 3}$O$_{\rm 7}$ (fig.1): at this level we expect the
occupied orbital at each V site to be d$_{xz}$ or d$_{yz}$ --- namely
the one parallel to the edge of the nn empty plaquette. 
The inclusion of more distant neighbours leads to a tilting of the principal 
axes of the EFG tensor, and consequently of the occupied orbital, 
from the x and y directions; however, for 
a cluster with 20 empty plaquettes, we obtained just a small tilting angle 
$\alpha \simeq$9$^{\circ}$ (see fig.2), the stabilization energy 
being $\Delta_{\rm CF} \simeq $0.18 eV. With this orbital structure the strongest 
antiferromagnetic interaction is that within dimers shown in fig.2, the next
one is J$_{\rm corner}$, while the edge exchange between dimers is weaker and
ferromagnetic. We thus expect
the 2-d lattice to be covered by singlet dimers with, because of the tilting, 
J$_{\rm dimers}$ somewhat smaller than J$_{\rm 1d}
(\simeq$ 150 K) estimated above for CaV$_{\rm 3}$O$_{\rm 7}$, in agreement with 
the observed spin gap of 107 K~\cite{Taniguchi95}.
These dimers can be additionally stabilized by the outward shifts of the V ions 
away from the points marked in fig.2 by triangles (the centers of the V
plaquettes in the "plaquette RVB" picture): it should be caused 
both by
the electric field, which we calculated to be about 1.1 10$^{\rm 6}$ in cgs units 
along the arrows at each V site, and by the dimer energy gain itself. These 
shifts are indeed experimentally observed in~\cite{Bouloux473}; note that they 
are opposite to those suggested by Starykh et al~\cite{Ueda96} within the 
plaquette RVB picture.\\
\hspace*{3mm} Finally we discuss the situation in CaV$_{\rm 2}$O$_{\rm 5}$ (fig.3), 
which 
turns out to be somewhat more delicate. The first striking feature of this
compound, usually treated as a spin ladder~\cite{Iwase96}, is the 
large value of the spin gap $\Delta_{\rm SG}$=616 K, even
larger than in copper ladders (typically $\sim$400 K~\cite{Dagotto96}). If we assume 
relatively isotropic interactions with J$_{\rm rung} \simeq$J$_{\rm leg}$,
we would need J$\geq$1200 K~\cite{Dagotto96}, i.e. of the same order as in 
High-T$_{\rm C}$
cuprates. But in contrast to the Cu$^{\rm 2+}$ case, for V$^{\rm 4+}$ 
we are dealing not with
e$_{\rm g}$ but with t$_{\rm 2g}$ electrons, and their much smaller 
overlap with the oxygen ligands should give a weaker exchange. We show
below that the possible solution of this puzzle lies again in the 
orbital ordering with the concomitant shift of the V ions and with the 
formation of tightly bound rung dimers.
The calculation of the EFG tensor at the
centre of each oxygen plaquette, where the V ions are supposed to be, shows
that the orbital d$_{\rm x'z}$, tilted 
by 45$^{\circ}$ with respect to the original x axis, would be there more
stable. However, the stabilization energy is in this case only
$\Delta_{\rm CF} \simeq$0.06 eV. With this type of orbital occupation, 
the large value of the 
spin gap~\cite{Iwase96} could hardly be explained, and no gap might exist at all. 
Straightforward calculation shows, however, that in the centre of the 
oxygen plaquettes there exists strong electric field along the rungs 
(2.1$\cdot$10$^{\rm 6}$ in cgs units). Consequently, we expect a shift 
of the V ions in the 
direction of the arrows in fig.3, which shortens the V-V distances in the 
rungs. If this shift $\delta$ is taken into account in the calculation of the 
EFG tensor, the splitting $\Delta_{\rm CF}$ changes sign for 
$\delta >\delta_{\rm C}\simeq0.2\AA$. For larger shifts the different orbital, 
d$_{\rm y'z}$, lies lower in energy, its stabilization energy rapidly increasing
with $\delta$: the resulting orbital ordering
is shown in fig.3. Such a structure can
explain the experimental observations: we would have here a strong A-A'
interaction (180$^{\circ} \pi$-bonding), and much weaker A-B and A-C
interactions, so that the picture would again predominantly correspond to
weakly coupled dimers. Note that here J$_{\rm AB} < $J$_{\rm AC}$, and the ladder 
model is thus completely inapplicable. In order to explain the
spin gap of 616 K~\cite{Iwase96}, we need in this framework 
J$_{\rm dimer}$ of the same order, i.e. $\sim$4 times 
stronger than the $\sim$150 K we estimated for 
CaV$_{\rm 3}$O$_{\rm 7}$. Several factors actually strenghten this interaction:
the in-plane shift of the V ions decreases the V-V 
and V-bridging oxygen
distances enhancing the overlap; moreover, we have here 
180$^{\circ}$-exchange in contrast to the weaker 90$^{\circ}$-exchange in 
chains in  
CaV$_{\rm 3}$O$_{\rm 7}$ and in dimers in CaV$_{\rm 4}$O$_{\rm 9}$ (note
however that there are two oxygens contributing to the exchange in the
latter cases). The main factor however is the out of plane position of the V ions.
This leads to an extra 
contribution to the exchange due 
not only to $\pi$- but also to $\sigma$-overlap (fig.4),
which depends on the angle $\gamma$ between the direction 
V-planar oxygen and the basal plane,
 $\tilde{t}_{\rm pd \sigma}=
t_{\rm pd \sigma}cos\gamma \sin(2\gamma)$. The d-p $\sigma$ hybridization 
t$_{\rm pd \sigma}$
is typically $\sim$2 times stronger than the $\pi$ one~\cite{Harrison}. As the
exchange interaction J$\sim t_{\rm pd}^{\rm 4}$, 
$\sigma$-hybridization gives significant contribution 
even for small $\gamma$, being much smaller for 90$^{\circ}$-exchange in 
CaV$_{\rm 3}$O$_{\rm 7}$ (where it is also present) due to partial
cancellation of different terms.
Combining all the contributions to the exchange, we get for the real 
structure ($\gamma\simeq$25$^{\circ}$) that the total interaction in dimers 
in CaV$_{\rm 2}$O$_{\rm 5}$ is about 
three times stronger than
in chains of CaV$_{\rm 3}$O$_{\rm 7}$ even without taking into account
shifts of the V ions. Thus this 
orbital ordering
can explain the properties of CaV$_{\rm 2}$O$_{\rm 5}$, although more detailed 
experiments and
calculations are needed to check the proposed picture.
Summarizing, we have shown that the properties of the
system CaV$_{\rm n}$O$_{\rm 2n+1}$, often treated as electron counterpart of the 
Cu compounds, can be easily 
understood with the orbital degrees of freedom taken into account: 
an orbital ordering should exist 
in these vanadates, which drastically modifies their 
exchange interactions and strongly influences their magnetic properties.
The type of ordering obtained for CaV$_{\rm 3}$O$_{\rm 7}$ makes it a quasi 1-d
antiferromagnet and
quite naturally explains its unusual magnetic structure, without the 
necessity to invoke complicated quantum effects. Ordinary spin singlet dimers 
should be formed in 
CaV$_{\rm 4}$O$_{\rm 9}$ as a result of the orbital ordering, instead of the 
plaquette RVB 
previously assumed. The most natural explanation of the properties of 
CaV$_{\rm 2}$O$_{\rm 5}$ with its unexpectedly large spin gap is the formation 
of tightly bound dimers
due to V ions shift and corresponding orbital ordering. Thus, the materials of 
the series 
CaV$_{\rm n}$O$_{\rm 2n+1}$ provide yet another example of how the
orbital structure of transition metal compounds~\cite{Goodenough,Kugel73,Pen96} 
often makes 
the homogeneus Heisenberg exchange inapplicable, giving rise to peculiar 
magnetic structures and
nontrivial magnetic properties. Finally, we point out that the
crystal structure of CaV$_{\rm 4}$O$_{\rm 9}$ is chiral. To our knowledge,
this is the first example of structurally chiral magnetic material. The unique 
combination 
of this chirality
(and corresponding spontaneus optical activity) with the presence of localized 
magnetic moments,
opens a broad spectrum of new possibilities and can lead to interesting physical 
consequences, e.g. in magnetooptical
properties etc.\\
\hspace*{3mm} We express our gratitude to G. Sawatzky and W. Geertsma for very useful 
discussions. This
work was supported by the Netherland Foundation for the Fundamental Study
of Matter (FOM).

\begin{figure}
\caption{CaV$_{\rm 3}$O$_{\rm 7}$: crystal and magnetic structure.
Oxygen ions are at the corners of the plaquettes, and V ions in the center.
 Orbital ordering is also shown (see the text).} 
\end{figure}

\begin{figure}
\caption{CaV$_{\rm 4}$O$_{\rm 9}$: crystal structure and orbital ordering. Singlet 
dimers are
marked by dashed ovals.}
\end{figure} 

\begin{figure}
\caption{CaV$_{\rm 2}$O$_{\rm 5}$: crystal structure and 
 possible orbital ordering. The usual "ladder" is shown.}
\end{figure}

\begin{figure}
\caption{CaV$_{\rm 2}$O$_{\rm 5}$: the extra $\sigma$-overlap.}
\end{figure}

\end{document}